\documentclass{article}

\usepackage{arxiv}

\usepackage[utf8]{inputenc} 
\usepackage[T1]{fontenc}    
\usepackage{hyperref}       
\usepackage{url}            
\usepackage{booktabs}       
\usepackage{amsfonts}       
\usepackage{nicefrac}       
\usepackage{microtype}      
\usepackage{graphicx}
\usepackage[numbers]{natbib}
\usepackage{doi}
\usepackage{amsmath}

\title{Well-to-Tank Carbon Intensity Variability of Fossil Marine Fuels: \\ A Country-Level Assessment}

\author{
    Wennan Long\textsuperscript{1}, Diego Moya\textsuperscript{2}, Zemin Eitan Liu\textsuperscript{1}, Zhenlin Chen\textsuperscript{3}, Liang Jing\textsuperscript{2}, \\
    \textbf{Muhammad Yousuf Jabbar\textsuperscript{2}}, 
    \textbf{Dimitrios Orfanidis\textsuperscript{4}}, \textbf{Mohammad S. Masnadi\textsuperscript{1*}} \\
    \textsuperscript{1}Department of Chemical and Petroleum Engineering, University of Pittsburgh, Pittsburgh, PA 15261, USA.\\
    \textsuperscript{2}Energy Traceability Technology, Technology Strategy and Planning Dept, Saudi Aramco, Kingdom of Saudi Arabia.\\
    \textsuperscript{3}Department of Energy Science and Engineering, Stanford University, Stanford, CA 94305, USA.\\
    \textsuperscript{4}Downstream Investment Analysis, Downstream Strategy and Capital Planning Dept, Saudi Aramco, Kingdom of Saudi Arabia.\\
    \textsuperscript{*}Corresponding author. Email: m.masnadi@pitt.edu
}

\renewcommand{\shorttitle}{Well-to-Tank Country-Level Marine Fuel CI}

\hypersetup{
pdftitle={Global Well-to-Tank Carbon Intensity Variability of Hydrocarbon-Based Marine Fuels},
pdfsubject={Marine Fuels, Carbon Intensity, HSFO, LPG, Policy},
pdfauthor={John Doe, Jane Smith},
pdfkeywords={Marine Fuels, Carbon Intensity, HSFO, LPG, Well-to-Tank, Policy},
}

\begin{document}
\maketitle

\begin{abstract}
The transition toward a low‐carbon future in maritime transportation requires a detailed understanding of the lifecycle carbon intensity (CI) of marine fuels. Irrespective of the feedstock, all well-to-tank (WTT) emissions contribute significantly to the sector's total greenhouse gas emissions; yet many studies lack a global perspective and only partially account for upstream operations, crude oil transportation, refining, liquids shipping and distribution. This study evaluates the WTT CI of High Sulphur Fuel Oil (HSFO) and well-to-refinery exit (WTR) CI of Liquefied Petroleum Gas (LPG) on a worldwide scale at asset level. HSFO represents a traditional, widely used marine fuel, while LPG serves as a potential transition fuel due to its lower tank-to-wake emissions and compatibility with low-carbon fuels such as ammonia. Using the Oil Production Greenhouse Gas Emissions Estimator (OPGEE) and Petroleum Refinery Life Cycle Inventory Model (PRELIM) tools, along with R-based geospatial and statistical methods, this work derives country-level CI values for 72 countries (HSFO) and 74 countries (LPG), covering 98\% of global HSFO and LPG refinery production. Results show that marine fuels produced around the world exhibit significantly different climate impacts, underscoring that not all fuels are created equal. HSFO upstream CI ranges from 1 to 22.7 gCO$_2$e/MJ, refining CI from 1.2 to 12.6 gCO$_2$e/MJ, and a global volume-weighted-average WTT CI of 12.4 gCO$_2$e/MJ. For HSFO, upstream and refining account for 55\% and 32\% of the WTT CI, respectively, with large-scale exporters and intensive refining practices (e.g., Russia, China, United States, Iran) having higher emissions. In refinery-sourced LPG pathways defined by the WTR boundary, upstream CI ranges from 0.9 to 22.7 gCO$_2$e/MJ, refining CI ranges from 2.8 to 13.9 gCO$_2$e/MJ, and the volume-weighted-average WTR CI is 15.6 gCO$_2$e/MJ. Refining makes up 49\% of the total LPG WTR CI, while upstream and transport represent 44\% and 6\%, respectively. Major players of the LPG sector include China, the United States and Russia. These findings reveal significant WTT and WTR CIs variability across countries and supply chains, offering opportunities for targeted policies to reduce emissions. \end{abstract}


\section{Introduction}
\label{sec:intro}

The maritime sector plays a central role in global trade, with about 90\% of global goods transported by sea \cite{MarineBenchmark2020}. In 2023, the total global consumption of fossil marine fuels reached 5.1 million barrels per day (b/d). In the same year, fossil marine fuels indicated a demand of approximately 1.1 million b/d of Heavy Sulfur Fuel Oil (HSFO), 1.5 million b/d of marine gasoil, and 2.5 million b/d of Very Low Sulfur Fuel Oil (VLSFO) \cite{WoodMackenzie2024}. Global fossil marine fuel consumption is expected to reach a peak in the mid-2020s at just over 5.3 million b/d. Meanwhile, global fossil marine fuel sales are forecast to grow only 2\% between now and 2030; significantly less than the growth observed in international maritime trade, mainly driven by efficiency gains and the adoption of alternative fuels. These trends underscore the importance of improving fuel quality and reducing emissions in the shipping industry and the potential use of Liquefied Petroleum Gas (LPG), marine Liquefied Natural Gas (LNG), biofuels and e-fuels.

HSFO, a residual refinery product containing various impurities, remains in use for vessels equipped with exhaust gas cleaning systems (scrubbers). Following the International Maritime Organization’s (IMO) 2020 regulation limiting sulfur content in marine fuels to 0.5\%, VLSFO has emerged as the main fuel in maritime applications. Yet, HSFO still accounts for a considerable portion of shipping fuel, even though globally produced HSFO is estimated at about 6.7 million b/d, with 16--17\% (around 1.1 million b/d) consumed by the marine sector. Meanwhile, alternative fuels, such as LNG, are gaining traction for their lower sulfur and carbon footprints. In contrast, LPG sees limited adoption despite its potential for a lower carbon footprint when sourced from non-refinery processes. \cite{Xydas2023}. Reliable, globally high resolution data on LPG use in the marine sector is limited. Corporate estimates suggest that LPG currently accounts for less than 2\% of total marine fuel consumption. This relatively modest share reflects the fact that LPG remains a niche fuel in shipping. However, its uptake is constrained by factors such as limited bunkering infrastructure and engine compatibility, while conventional fuels (e.g., HSFO, LSFO/VLSFO and marine gasoil) dominate the market. As interest in low-carbon alternatives grows and pilot programs for LPG bunkering expand, more consistent data on its use may emerge.

In the 2023 IMO Greenhouse Gas (GHG) Strategy, which extends the framework set out in 2018, the organization aims to achieve net-zero GHG emissions by or around 2050. The strategy also includes checkpoints of reducing total annual GHG emissions by 20\% (striving for 30\%) by 2030 and 70\% (striving for 80\%) by 2040, both against a 2008 baseline \cite{IMO2023a, IMO2022a, IMO2022b, IMO2022c, IMO2022d}. Various studies have investigated the environmental impacts of marine fuels. Bengtsson et al. \cite{Bengtsson2012} performed an early life cycle assessment (LCA) comparing Heavy Fuel Oil (HFO), Marine Gas Oil (MGO), Liquefied Natural Gas (LNG), and Gas to Liquids (GTL) under a consequential framework, while Gilbert et al. \cite{GILBERT2018855} considered HFO, Marine Diesel Oil (MDO), LNG, hydrogen, methanol, and biofuels using an attributional model. More recently, El-Houjeiri et al. \cite{ElHoujeiri2018} studied HFO, MGO, and LNG supply chains spanning Saudi Arabia, the North Sea, Australia, Qatar, and the United States, and Chen and Lam \cite{CHEN2022103192} compared diesel and hydrogen with data from the European Reference Life Cycle Database (ELCD) and Ecoinvent 3.6. 

The reviewed literature consistently demonstrates that the estimated carbon intensities (CI) of marine fuels vary widely across studies due to differences in upstream operations, feedstock quality, refining processes, operational practices, and the chosen system boundaries. This variability not only highlights the inherent complexity in accurately quantifying lifecycle emissions but also underscores the heterogeneous challenges in establishing a universal benchmark for marine fuel performance. As the IMO's 2023 Greenhouse Gas Strategy sets ambitious targets toward Net-Zero emissions by 2050, this CI variability calls for standardized methodologies and more robust, transparent data across the entire marine fuel supply chain, ultimately informing more effective policies and accelerating the transition to low-carbon shipping solutions.

Studies have increasingly documented how well-to-tank (WTT) CI diverge across regions, as shown in Table~\ref{tab:fuel_ci_regions}. Fuel type, refining processes, transport distances, and local energy inputs all affect carbon intensity. Nonetheless, many analyses focus on specific fuels or geographies without providing country-level insights into conventional pathways of HSFO and refinery-sourced LPG production. Lifecycle emissions remain a central concern in shipping research. Jing et al. \cite{jing2022understanding} reported jet fuel well-to-wake (WTW) CI ranging from 81.1 to 94.8\,gCO$_2$e/MJ, while Ha et al. \cite{ha2023framework} linked large variations in marine fuels to crude source differences. Kourkoumpas et al. \cite{kourkoumpas2024life} further highlighted emerging production methods for aviation and marine fuels that could reduce emissions in the next decades.

\begin{table}[ht!]
  \centering
  \caption{Well-to-Tank CI of Various Fuel Types Across Different Regions}
  \label{tab:fuel_ci_regions}
  \begin{tabular}{lll l}
    \toprule
    Fuel Type & Well-to-Tank CI (g CO$_2$eq/MJ) & Study Region & Source \\
    \midrule
    Heavy Fuel Oil (HFO, 2.7\% S) & 10--15 & Global & Hawkins et al. (2023) \cite{Hawkins2019} \\
    Marine Gas Oil (MGO, 1.0\% S) & 8.0--12 & Global & Hawkins et al. (2023) \cite{Hawkins2019} \\
    Liquefied Natural Gas (LNG) & 2.3--6.0 (varies by import distance) & South Korea & Ha et al. (2023) \cite{HA2023165366} \\
    Biodiesel & 10--20 (feedstock dependent) & USA & Hawkins et al. (2023) \cite{Hawkins2019} \\
    Blue Ammonia (SMR + CCS) & $\sim$30 & Global & Seddiek \& Ammar (2023) \cite{SEDDIEK2023103547} \\
    Green Ammonia (Electrolysis) & 0.0--5.0 (depends on electricity source) & Global & Seddiek \& Ammar (2023) \cite{SEDDIEK2023103547} \\
    Hydrogen (Green, Electrolysis) & 0.0--5.0 (depends on electricity source) & Singapore & Chen \& Lam (2022) \cite{CHEN2022103192} \\
    Bio-Methanol & 5.0--20 & Europe & Strazza et al. (2010) \cite{STRAZZA20101670} \\
    \bottomrule
  \end{tabular}
\end{table}

Despite significant efforts by the scientific community to quantify the carbon footprint of fossil marine fuels and address the challenges of benchmarking across countries and assets, global institutions such as the IMO and IPCC continue to recommend life cycle GHG intensities for marine fuels that adopt single representative homogeneous values. The 2024 IMO LCA guidelines' approach \cite{IMO2024} applies default values for WTT, tank-to-wake (TTW), and WTW CIs across 14 maritime fuels, a practice that invites scientific criticism due to its oversimplification. For instance, for heavy fuel oil (HFO) with \(0.10 < S \leq 0.50\%\), the WTT CI is set at 16.8 gCO\(_2\)e/MJ, while for HFO exceeding 0.50\% S, it is 14.1 gCO\(_2\)e/MJ. For LNG, a conversion factor of 57.3 gCO\(_2\)/MJ (estimated from 2.750 gCO\(_2\)/g of fuel at an LCV of 0.0480 MJ/g) is given. However, these values are given without a detailed WTW assessment and without differentiation among processes, assets, or countries. Similarly, for LPG (propane), the guidelines imply a conversion factor of 64.8 gCO\(_2\)/MJ, derived from 3 gCO\(_2\)/g fuel and an LCV of 0.0463 MJ/g; yet do not specify a complete WTW CI. These values have remained similar to those in the 2023 IMO LCA guidelines' approach \cite{IMO2023b}. Although LPG is sometimes considered a transitional or alternative option in shipping, its actual uptake remains small. Moreover, approximately two thirds of LPG worldwide is derived from natural gas liquids (NGLs), which may exhibit lower carbon intensities than refinery-sourced LPG. The lack of consistent data on non-refinery LPG entering marine usage, combined with limited regulatory assessments that assume a single average emission factor for LPG, underscores the need for more refined system boundaries, emission assessments, and data inputs before drawing conclusions on LPG’s role in decarbonizing shipping.

The study here addresses a significant gap in the current literature on the LCA of fossil marine fuels. Although many studies quantify the carbon footprint of fossil marine fuels within narrowly defined boundaries and often fail to capture the full LCA, reputable global institutions continue to recommend using single, homogeneous default values for CI or conversion factors. This practice, which oversimplifies the inherent complexity and heterogeneity of fuel production and use, may misinform decision makers, investors, and policymakers regarding decarbonization strategies of the marine sector. In response, this study aims to assess two well‐established fuels: High Sulphur Fuel Oil, currently used in this sector, and refinery-sourced Liquefied Petroleum Gas, a potential alternative. The study evaluates these fuels at the asset level, using a WTT boundary (upstream, transportation, refining, shipping and distribution) for HSFO and a WTR boundary (upstream, transportation and refining) for LPG, with results presented at the country level. The novelty of our work lies in its bottom-up, engineering-rich LCA approach, which incorporates data from approximately 9,000 oilfields across 93 countries, 500 refineries covering 98\% of global crude oil and fuel production, 3,500 crude-to-refinery links, and over 8,000 HSFO trading routes, thereby yielding high-resolution CI estimates at refinery, country, and route levels. To achieve this, the study applies the Oil Production Greenhouse Gas Emissions Estimator (OPGEE) to quantify upstream emissions and the Petroleum Refinery Life Cycle Inventory Model (PRELIM) to estimate refinery-level emissions, complemented by advanced geospatial big data analytics.

\section{Methodology}
\label{sec:methodology}

\begin{figure}[ht!]
    \centering
    \includegraphics[width=0.8\textwidth]{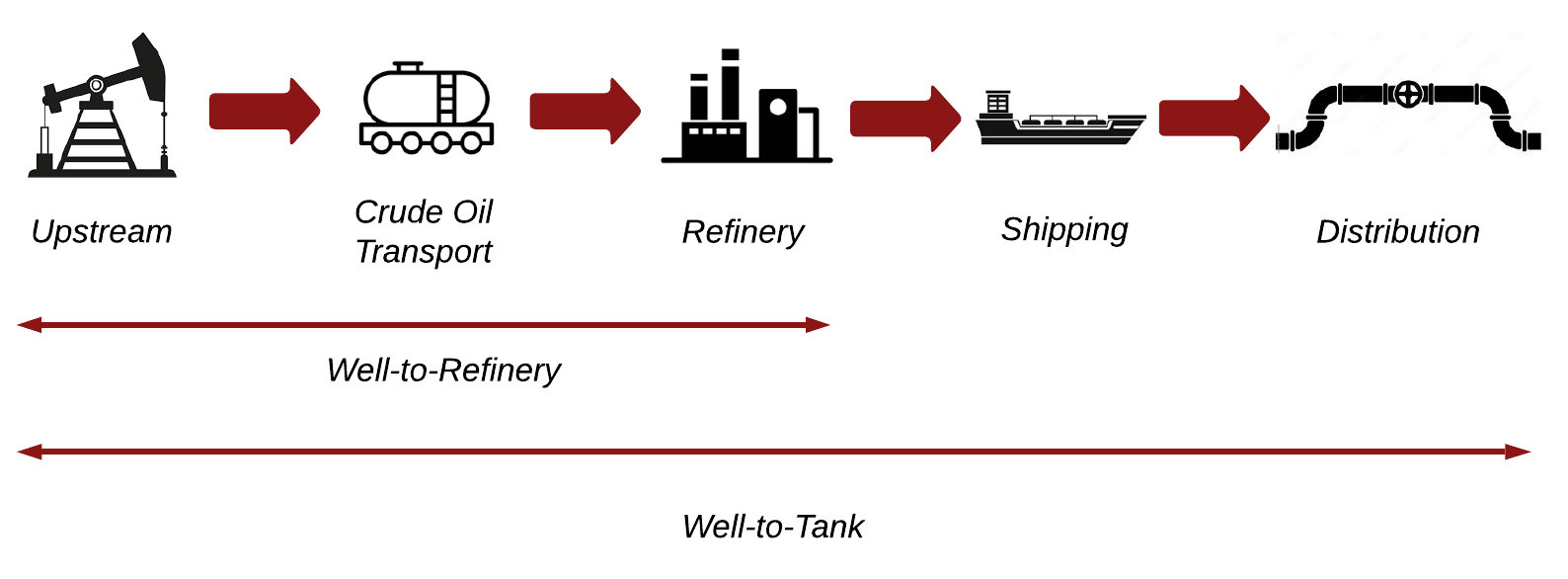}
    \caption{System boundaries for HSFO (well to tank) and LPG (well to refinery) carbon intensity analyses. HSFO includes upstream (exploration, drilling, and crude oil production), transport, refining, shipping, and distribution, whereas LPG concludes at the refinery exit gate. Fuel combustion is outside the scope of this paper.}
    \label{fig:system_boundary}
\end{figure}

This study evaluates carbon intensity (CI) in two distinct system boundaries for marine fuels. For High Sulfur Fuel Oil (HSFO), the boundary extends from the well to the tank (WTT), while for Liquefied Petroleum Gas (LPG) it concludes at the refinery exit gate (well to refinery, WTR). Figure~\ref{fig:system_boundary} illustrates these boundaries, which begin with upstream processes (exploration and crude oil production), continue through crude oil transport and refining, and, in the case of HSFO, also include shipping and distribution. The analysis excludes combustion, ensuring results end at the point of fuel readiness for maritime use. Due to limited data on how non-refinery LPG is traded and consumed in shipping, the scope here focuses solely on refinery-sourced LPG and excludes shipping emissions for LPG to avoid conflating non-refinery-based volumes.

\subsection{Data Source and Data Processing}
Marine fuels’ life cycle greenhouse gas (GHG) emissions are quantified using a process-based attributional life cycle assessment (LCA) approach that systematically accounts for mass and energy flows along the entire supply chain (e.g., WTT for HSFO and WTR for LPG). Crude oil production emissions, including both extraction and processing, are sourced from the seminal work of \cite{masnadi2018Science} and \cite{Ankathi2024MENA-GREET}. Additionally, emissions from crude oil transportation via pipelines are derived from the model presented by \cite{Dixit2023NNCC}, which employs the Crude Oil Pipeline Transportation Emissions Model (COPTEM) to accurately calculate pipeline-related GHG emissions.

Emissions from refining crude oil into marine fuels are estimated using PRELIM v1.6. Detailed data on crude oil assays and refinery operations are obtained from proprietary commercial datasets, including Wood Mackenzie, that are not available in the public domain. These crude assays are converted into a PRELIM-compatible format that segments the crude oil into nine distinct temperature fractions, from light straight run to vacuum residue, thereby ensuring precise emissions attribution across refining cuts. Furthermore, Heavy Sulfur Fuel Oil (HSFO) trade data are sourced from subscription-based commercial databases, including S\&P Global, while shipping emissions associated with LPG trade are calculated based on the methodology developed by \cite{ha2023framework}.

\subsection{OPGEE}
OPGEE is an engineering-based life cycle assessment model developed to calculate upstream greenhouse gas (GHG) emissions from oil and gas operations \cite{el2012opgee, el2013lca, el2017opgee, masnadi2019opgee}. Since its initial release, the model has expanded to include detailed equipment- and process-level representations of flaring, enhanced oil recovery, and other key factors influencing CI. This study uses the most recent Python-based implementation, covering exploration, drilling, and crude oil production stages.

Field-level input data is obtained from both commercial and publicly available datasets. Commercial databases offer detailed field and well parameters under a subscription model, while public sources provide additional regional details. For instance, Canadian oil sands data from the Alberta Energy Regulator inform specialized cases with unique steam-oil ratios. Other research datasets address gaps in parameters such as water-oil ratios or natural gas reinjection fractions. A Python-based pipeline merges and reconciles these inputs by applying a hierarchical approach: missing values are filled using regionally averaged or default estimates, and higher-priority datasets override lower-priority information. Conflicts, such as fields identified as using water flooding without a corresponding water injection ratio, are systematically resolved by correcting production methods or adjusting the relevant parameters based on corroborating data.

Where reservoir depth is unavailable, the model applies generic correlations to approximate reservoir pressure and temperature. Similarly, missing well counts may be inferred from oil production rates using an empirical ratio that accounts for variations in artificial lift or water injection. Associated gas flaring is estimated through country-level flaring oil ratios, which draw on satellite observations to derive total flared gas volumes under assumed flaring efficiencies. In cases with multiple data sources, the pipeline assigns priority to verified or newer information, ensuring alignment with actual field practices. Parent-child relationships—common in large reservoirs partitioned into multiple operating blocks—are also reconciled, with high-level production methods cascaded to local units unless more precise data exists.

\subsection{PRELIM}
PRELIM v1.6 is an advanced, open-source, data-driven methodology that quantifies energy use and greenhouse gas emissions in petroleum refineries at a detailed process-unit level. Developed at the University of Calgary, PRELIM adopts a systems-level approach by integrating rigorous mass and energy balance calculations with empirical correlations from both public data and proprietary industry datasets \cite{PRELIM2025}. Up to seven distinct refinery configurations can be simulated, ranging from basic hydroskimming to complex deep-conversion systems, drawing on detailed crude assay data (including distillation curves, API gravity, sulfur, and hydrogen content) to partition the crude into multiple fractions. Through linear programming and a flexible allocation framework, which distributes energy inputs and emissions based on mass, energy content, market value, or hydrogen content, PRELIM enables transparent, reproducible estimates of process-specific energy consumption and associated environmental impacts \cite{Jing2020}. In addition, by incorporating the U.S. EPA’s TRACI life cycle impact assessment methodology, PRELIM extends beyond energy accounting to evaluate multiple environmental impact categories, making it a versatile resource for academic research and policy analysis.

\subsection{Shipping LCA}
An integrated method and algorithm are applied to estimate the carbon intensity (CI) of shipping heavy fuel oil (HSFO) using detailed liquid bulk cargo trade data from commercial corporate datasets. The process begins by standardizing variables in the cargo movement dataset, then using custom mapping functions—enhanced with geospatial coordinates from the \texttt{rnaturalearth} package in R—to determine average distances between loading and discharging countries. These distances, computed via the Vincenty formula in the \texttt{geosphere} package, yield precise travel estimates and validate a vessel’s final discharging port and its last port-of-call prior to loading. Next, shipping CI values (kg\,CO$_2$ per barrel) are assigned from a reference dataset for known country pairs, based on shipping distances reported in \cite{jing2022understanding}. For routes without direct matches, an iterative binning approach groups similar distances and imputes average CI values, using a nearest-neighbor algorithm tailored for the shipping data. The algorithm then incorporates refinery CI data by merging well-to-refinery metrics \cite{Dixit2023, Jing2020} for each producing country, allowing distinctions between domestic fuel production and trade imports or exports. Finally, the methodology aggregates trade volumes and computes volume-weighted average CI at the refinery, route, and country levels, generating summary tables that support further analysis of HSFO shipping emissions. This comprehensive approach also allows net HSFO volumes to be estimated for each country, accounting for both domestic production and cross-border trade. 

For refinery-based LPG, shipping emissions are not included here because most LNG carriers and multi-gas carriers transport a mix of NGL-derived LPG and refinery-sourced LPG. Attributing all transport emissions to refinery-sourced volumes alone would overstate the shipping intensity for that pathway, given that a significant share of LPG originates outside of refineries. Consequently, this approach reports emissions up to the refinery exit gate for LPG, reflecting the recognized data gap in distinguishing non-refinery LPG flows.

\section{Results and Discussion}

\subsection{HSFO Well-to-Tank Carbon Intensity variability}
\begin{figure}[!ht]
    \centering
    \includegraphics[width=0.9\textwidth]{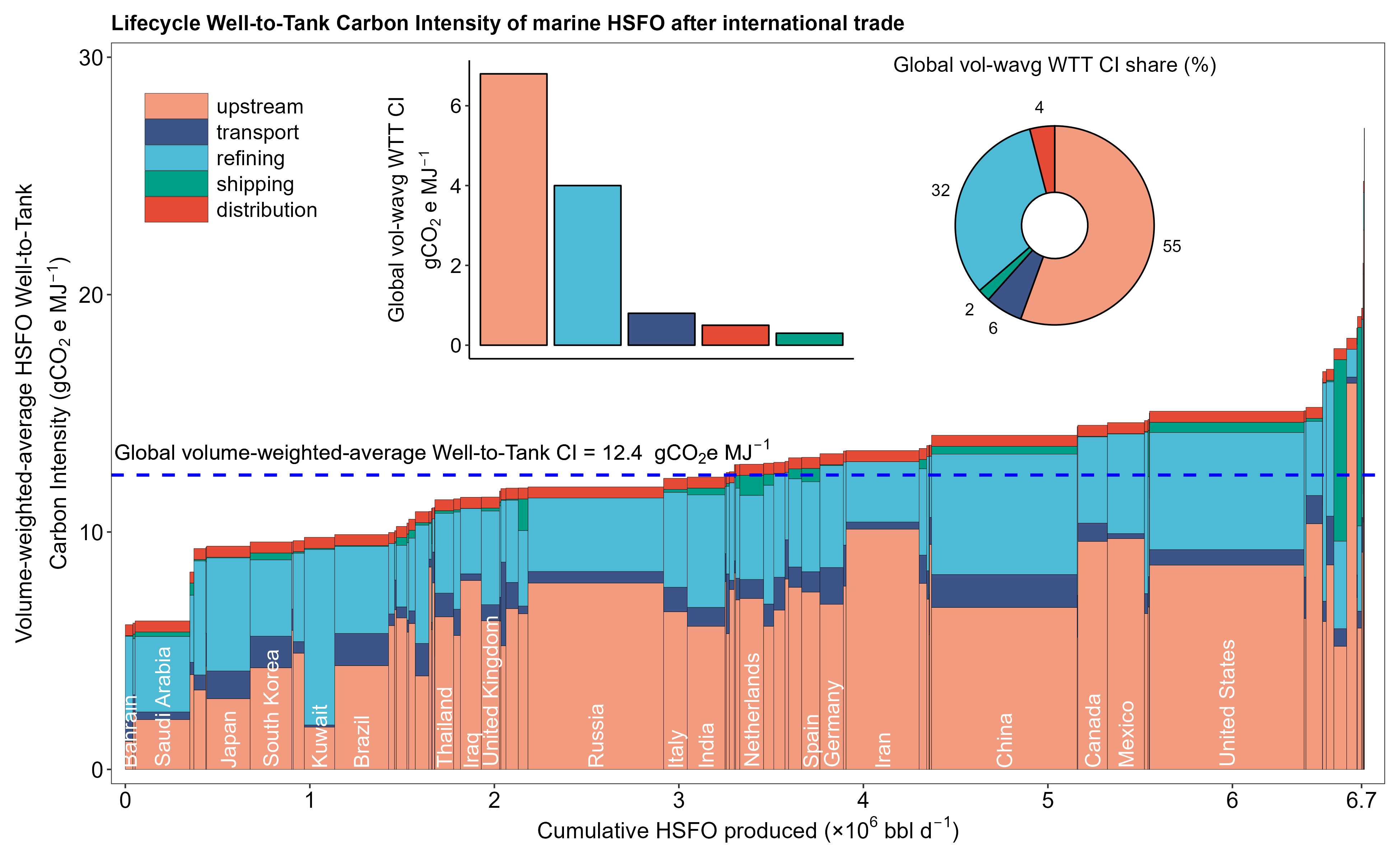}
    \caption{\textbf{Variable Lifecycle Well-to-Tank Carbon Intensity of Marine HSFO After International Trade.}
    The bar graph details WTT emissions across upstream, transport, refining, shipping, and distribution. The dashed line marks the global volume-weighted average (12.4\,gCO$_2$e/MJ). The inset bar chart shows the global WTT carbon intensity contributed by each segment, and the pie chart illustrates the percentage share of each segment in the overall average.}
    \label{fig:fig1}
\end{figure}

Figure~\ref{fig:fig1} displays a comparative analysis of the WTT carbon intensity for marine HSFO after international trade, covering a total HSFO production of about $6.7\times10^6$\,bbl/d. Approximately 20--25\% of this HSFO is consumed by maritime shipping, though precise country-level consumption volumes are not publicly available. Emissions are segmented by upstream operations, transport, refining, shipping, and distribution, with upstream (1--22.7\,gCO$_2$e/MJ) and refining (1.2--12.6\,gCO$_2$e/MJ) dominating the overall footprint. The inset bar chart indicates that upstream contributes the largest share at 55\%, followed by refining at 32\%, transport at 6\%, shipping at 4\%, and distribution at 2\%. Countries such as Saudi Arabia, South Korea, and Japan register relatively lower carbon intensities compared to major producers like Russia, the United States, and China, whose higher values stem from both large production volumes and varied upstream operations. Smaller-volume nations, including Serbia, Uzbekistan, Hungary, Austria, and Bulgaria, also exhibit elevated per-unit emissions. The global volume-weighted-average WTT CI is revised to 12.4\,gCO$_2$e/MJ, reflecting updated shipping volumes and domestic distribution. Flaring intensity, crude quality, and refinery efficiency remain the key explanatory factors, and modern refining techniques or cleaner upstream activities correlate with lower observed carbon intensities. These results clearly illustrate that HSFO carbon intensity varies significantly due to the complex and heterogeneous operations at the asset level, from well to final distribution.

\subsection{Refinery-Sourced LPG Well-to-Refinery Exit Gate Carbon Intensity variability}
\begin{figure}[!ht]
    \centering
    \includegraphics[width=0.9\textwidth]{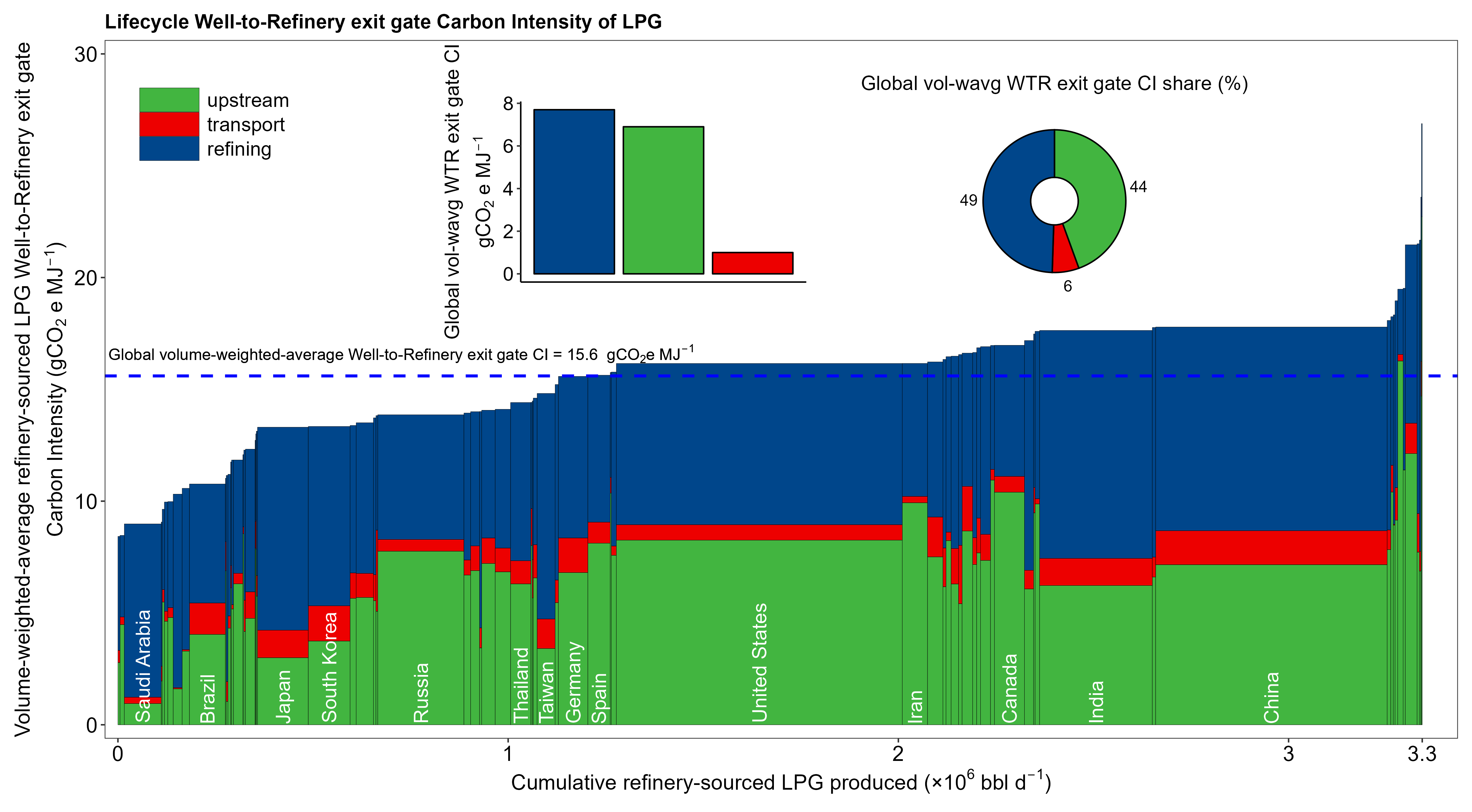}
    \caption{\textbf{Variable Lifecycle Well-to-Refinery Exit Gate Carbon Intensity of Refinery-Sourced LPG.}
    The graph displays upstream, transport, and refining emissions for refinery-sourced LPG at a global production rate of $3.3\times10^6$\,bbl/d. The dashed line represents the volume-weighted-average exit gate carbon intensity of 15.6\,gCO$_2$e/MJ. The inset bar chart indicates segment-specific intensities, and the pie chart illustrates the percentage contribution of each segment to the global average.}
    \label{fig:fig2}
\end{figure}

Figure~\ref{fig:fig2} presents a comparative analysis of refinery-sourced Liquefied Petroleum Gas (LPG) carbon intensity at the refinery exit gate, covering a total production of about $3.3\times10^6$\,bbl/d. Global statistics indicate that a substantial portion of total LPG (often over 60\%) derives from natural gas liquids (NGLs) rather than refineries, and it remains uncertain how much of this pool enters marine usage. Emissions here are thus reported only up to the refinery exit gate to avoid conflating transport for non-refinery LPG streams. The figure segments emissions into upstream (green), transport (red), and refining (blue). Refining constitutes the dominant share for most countries, reflecting additional processing steps required for LPG. The inset bar chart shows that refining accounts for 49\% of emissions, with upstream at 44\% and transport at 6\%. Countries like Saudi Arabia, South Korea, and Japan exhibit moderate intensities, while the United States and China show higher overall values due to large production volumes and varying refinery configurations. Smaller producers, including Iran and Taiwan, demonstrate diverse outcomes based on specific operating conditions. The global volume-weighted-average at the refinery exit gate is 15.6\,gCO$_2$e/MJ, indicating that both upstream operations and refining strategies significantly affect carbon intensity. Because shipping is excluded for refinery-sourced LPG, this figure offers a conservative estimate of potential well-to-tank impacts should the fuel be used in marine engines.

\subsection{Comparison of HSFO and LPG Pathways}
A comparison of HSFO and refinery-sourced LPG reveals overlapping carbon intensity ranges but differing key drivers. HSFO tends to show greater upstream variability tied to crude extraction and flaring, whereas LPG pathways rely more on intensive refining, with non-refinery NGL-based LPG generally having lower overall emissions. Best practices in flaring control, crude selection, and refinery optimization can mitigate carbon intensity in both fuels. Nonetheless, data gaps, especially regarding how much non-refinery LPG enters marine usage, complicate precise comparisons. Although emerging marine fuels such as marine LNG, ammonia and hydrogen continue to attract attention, short-term reductions can be achieved by improving conventional pathways and employing rigorous data tracking to capture accurate life cycle emissions.

\section{Conclusion}
\label{sec:conclusion}

This study shows that well-to-tank carbon intensity varies significantly among countries for both High Sulfur Fuel Oil (HSFO) and refinery-sourced Liquefied Petroleum Gas (LPG). Factors such as upstream flaring intensity, crude quality, and refinery efficiency account for much of this variation. A single default CI value does not accurately reflect real-world processes and can mask improvement opportunities. Granular, country-level data reveals a spectrum of emissions performance, underscoring the need for targeted efforts, such as flaring reduction, carbon capture, electrification, and optimized refining, to achieve immediate gains in reducing maritime emissions across fossil marine fuel supply chains.

Although this analysis focuses on refinery-based LPG and excludes shipping emissions for that pathway, it acknowledges that a major share of global LPG originates from natural gas liquids with potentially lower carbon intensity. Lack of data on how non-refinery LPG is traded, blended, and ultimately consumed by the marine sector remains a critical limitation. Similarly, the absence of reliable country-level end-use volumes restricts insight into tank-to-wake assessments. Future work should incorporate refined data on NGL-based LPG and actual marine consumption patterns to clarify the full range of well-to-wake CI variability of marine fuels. Addressing these gaps will be essential for policymakers and industry stakeholders aiming to align with the IMO’s decarbonization targets and accelerate the transition to lower-carbon shipping solutions.

Refinery decarbonization strategies, such as renewable-sourced electrification of process units, carbon capture, and low-carbon hydrogen integration, alongside process optimization through digitalization, waste heat recovery, and advanced catalyst technologies, offer a promising pathway to reduce the life-cycle greenhouse gas emissions of marine fuels. Looking ahead, potential innovations such as biomass coprocessing, methane pyrolysis for hydrogen production, process intensification, and the development of synthetic fuel pathways offer additional avenues for refinery decarbonization. Together, these integrated approaches and advances provide a transformative strategy to decarbonize refinery operations by significantly lowering refining carbon intensity and aligning the sector with global climate targets. By paving the way for a resilient future in both marine fuel production and shipping, these advances serve as critical levers in the energy transition. Continued innovation and cross-sector collaboration will be essential to overcome technical and economic challenges, ultimately reshaping marine fuel production and making a substantive contribution to global climate mitigation efforts.

In the case of high carbon intensities from upstream operations (e.g., refinery-sourced LPG), reducing emissions requires a comprehensive approach that targets both process efficiency and mitigation of fugitive emissions. By embracing advanced digital monitoring systems, predictive maintenance, and energy-efficient equipment, operators can significantly optimize production processes and lower overall energy consumption. Simultaneously, robust measures such as effective gas capture, minimized flaring and venting, and proactive detection and repair of methane leaks are essential to address fugitive emissions. Integrating these strategies not only reduces the environmental footprint of crude oil extraction but also sets a critical precedent to reduce emissions throughout the supply chain of the marine fuels industry, aligning upstream operations with global climate mitigation goals.

\bibliographystyle{unsrtnat}
\bibliography{template}  

\end{document}